# Pseudogap formation above the superconducting dome in iron-pnictides

T. Shimojima[1,2,*], T. Sonobe[1], W. Malaeb[3,4], K. Shinada[1], A. Chainani[5,6], S. Shin[2,3,4,5], T. Yoshida[4,7], S. Ideta[1], A. Fujimori[4,7], H. Kumigashira[8], K Ono[8], Y. Nakashima[9], H. Anzai[10], M. Arita[10], A. Ino[9], H. Namatame[10], M. Taniguchi[9,10], M. Nakajima[4,7,11], S. Uchida[4,7], Y. Tomioka[4,11], T.Ito[4,11], K. Kihou[4,11], C. H. Lee[4,11], A. Iyo[4,11], H. Eisaki[4,11], K. Ohgushi[3,4], S. Kasahara[12,13], T. Terashima[12], H. Ikeda[13], T. Shibauchi[13], Y. Matsuda[13], K. Ishizaka[1,2]

[1] *Department of Applied Physics, University of Tokyo, Bunkyo, Tokyo 113-8656, Japan.*
[2] *JST, CREST, Chiyoda, Tokyo 102-0075, Japan*
[3] *ISSP, University of Tokyo, Kashiwa 277-8581, Japan.*
[4] *JST, TRIP, Chiyoda, Tokyo 102-0075, Japan*
[5] *RIKEN SPring-8 Center, Sayo, Hyogo 679-5148, Japan*
[6] *Department of Physics, Tohoku University, Aramaki, Aoba-ku, Sendai 980-8578, Japan*
[7] *Department of Physics, University of Tokyo, Bunkyo, Tokyo 113-0033, Japan.*
[8] *KEK, Photon Factory, Tsukuba, Ibaraki 305-0801, Japan.*
[9] *Graduate School of Science, Hiroshima University, Higashi-Hiroshima 739-8526, Japan.*
[10] *Hiroshima Synchrotron Center, Hiroshima University, Higashi-Hiroshima 739-0046, Japan*
[11] *National Institute of Advanced Industrial Science and Technology, Tsukuba 305-8568, Japan.*
[12] *Research Center for Low Temperature and Materials Sciences, Kyoto University, Kyoto 606-8502, Japan.*
[13] *Department of Physics, Kyoto University, Kyoto 606-8502, Japan.*

**The nature of the pseudogap in high transition temperature (high-$T_c$) superconducting cuprates has been a major issue in condensed matter physics. It is still unclear whether the high-$T_c$ superconductivity can be universally associated with the pseudogap formation. Here we provide direct evidence of the existence of the pseudogap phase via angle-resolved photoemission spectroscopy in another family of high-$T_c$ superconductor, iron-pnictides. Our results reveal a composition dependent pseudogap formation in the multi-band electronic structure of $BaFe_2(As_{1-x}P_x)_2$. The pseudogap develops well above the magnetostructural transition for low $x$, persists above the nonmagnetic superconducting dome for optimal $x$ and is destroyed for $x \sim 0.6$, thus showing a notable similarity with cuprates. In addition, the pseudogap formation is accompanied by inequivalent energy shifts in xz/yz orbitals of iron atoms, indicative of a peculiar iron orbital ordering which breaks the four-fold rotational symmetry.**

The pseudogap (PG) observed in the normal state of the high-$T_c$ copper oxide superconductors remains a mysterious state of matter [1-3]. It has been attributed to several mechanisms such as a precursor pairing [4-6] and a novel form of spin/charge ordering [7-9]. Nearly a quarter-century after the discovery of high-$T_c$ superconductivity [10], the PG phase is still extensively debated in the literature and no consensus has been reached regarding its origin. In order to gain further insights into the relationship between the high-$T_c$ superconductivity and the PG, the exploration of the PG phase in other high-$T_c$ superconductors is highly desired.

Iron-pnictides [11] are another class of high-$T_c$ superconductors whose typical phase diagram is shown in Fig. 1(a). The parent compound shows stripe-type antiferromagnetic (AF) ordering at $T_N$ accompanying a lattice distortion from tetragonal to orthorhombic structure at $T_s$ [12]. In contrast to the parent cuprate which is an AF insulator, the parent pnictide is an AF metal. The electronic structure derived from multiple Fe 3d orbitals [13], consists typically of disconnected hole and electron Fermi surfaces (FSs), which undergo an electronic reconstruction across $T_N$ and $T_s$ [14, 15]. In addition to spin fluctuations derived from the nesting between the disconnected FSs [16, 17], orbital fluctuations are also a candidate for a driving force of electron pairing in iron-pnictides [18, 19]. The importance of the orbital degrees of freedom in the itinerant ground states attracted much attention in terms of orbital ordering, defined as an inequivalent electronic occupation of zx/yz orbitals [20-22]. Intensive researches suggested that iron-pnictides can be distinguished from copper-oxides based on its orbital multiplicity and itinerant magnetism. It has been, thus, an important issue whether the high-$T_c$ iron-pnictide family also exhibits a PG phase in common with cuprates.

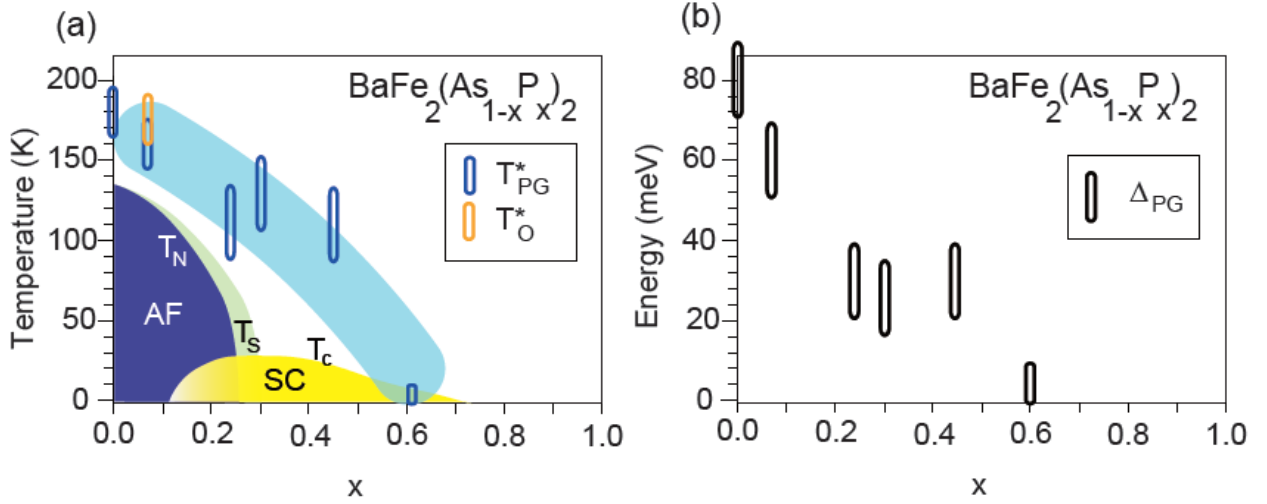

**Fig. 1.** (a) Phase diagram of BaFe$_2$(As,P)$_2$. Blue and orange markers indicate the temperatures of the PG formation $T^*_{PG}$ and inequivalent energy shift in zx/yz orbitals $T^*_O$, respectively. (b) Composition-dependence of the PG energy $\Delta_{PG}$ obtained by laser-ARPES.

One of the most intriguing properties of iron-pnictides recently reported is their two-fold rotational symmetry of various physical properties below $T_{N,s}$, as probed by transport [23], optical [24], scanning tunneling microscopy (STM) [25], inelastic neutron scattering [26] and ARPES measurements [14, 15]. There have been experimental reports showing the persistence of two-fold symmetries even above $T_{N,s}$ [15, 27-30]. In particular, the isovalent-substitution system BaFe$_2$(As$_{1-x}$P$_x$)$_2$ (AsP122) [31] shows an electronic nematic phase transition which breaks the rotational symmetry of the lattice. This transition occurs at a temperature $T^*_{Nem}$ well above $T_{N,s}$ and persists over the superconducting (SC) dome [27]. It should be noted that such a two-fold symmetric electronic nematic state has been reported in the PG region of high-$T_c$ cuprates [6].

Although several studies have been reported on the PG formation [32-40], the PG phase has not been completely established in the phase-diagrams of the iron-pnictides. In particular, only one angle-resolved photoemission spectroscopy (ARPES) study has shown the presence of a normal state energy gap in momentum space for Ba$_{0.75}$K$_{0.25}$Fe$_2$As$_2$ [41]. The overall picture of the PG formation in the multi-band electronic structure has not been clarified so far. In this work, we report the PG formation in the momentum-resolved electronic structure of AsP122 and its evolution with temperature and compositions, in pursuit of an as yet unknown PG phase.

Single crystals of BaFe$_2$As$_2$ were synthesized using the flux method. The starting materials Ba, FeAs were placed in an alumina crucible. This was then sealed in a double quartz tube under 0.3 atmosphere of Ar. The tube was heated at 1273 K for 2 h, slowly cooled to 1073 K for 24 h, and then quenched to room temperature ($T$) [42]. Single crystals of BaFe$_2$(As$_{1-x}$P$_x$)$_2$ with $x$ = 0.07, 0.24, 0.30, 0.35, 0.45, 0.61 were grown by a self-flux method [31,43]. The precursors of Ba$_2$As$_3$, Ba$_2$P$_3$, FeAs and FeP were mixed, and then sealed in a quartz tube. All the processes were carried out in a glove box filled with dry N$_2$ gas. The tube was heated at 1150 °C for 10 h, and slowly cooled down to 900 °C at a cooling rate 1 °C/h, followed by decanting the flux. The composition of the grown single crystals was confirmed by the energy dispersive X-ray analysis [43].

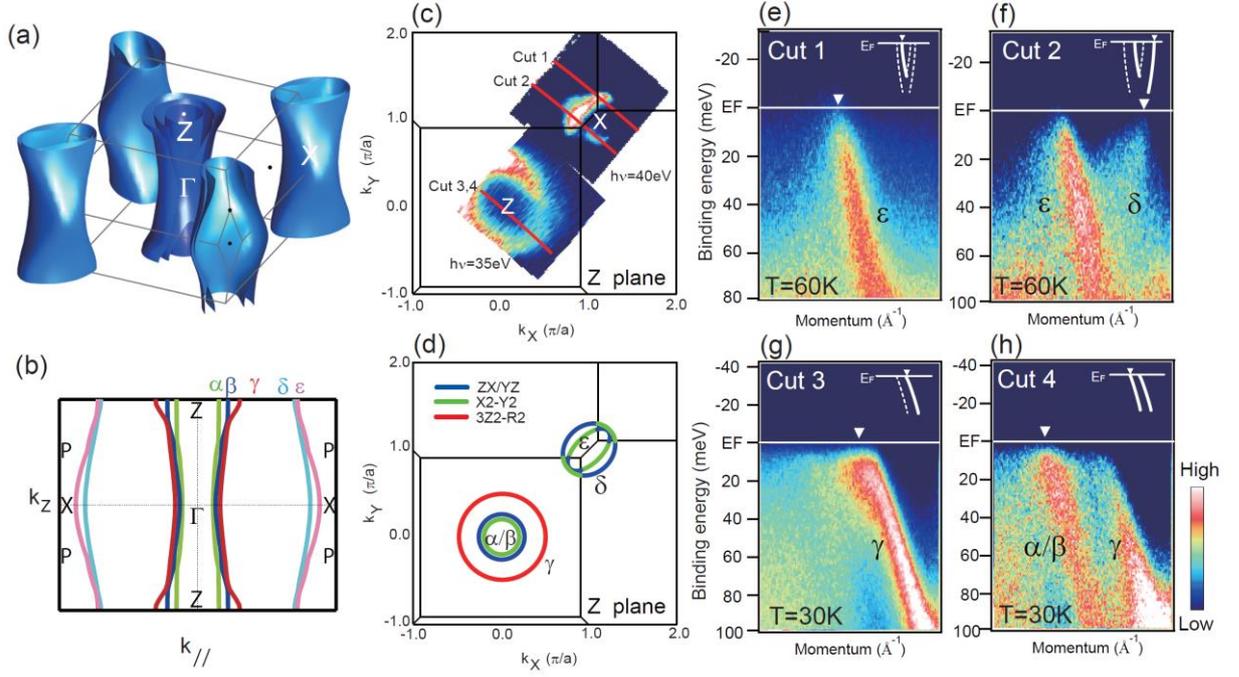

**Fig. 2.** (a) FSs of BaFe$_2$(As,P)$_2$ for $x = 0.30$ obtained by first principle band calculations. (b) Vertical cut of the calculated FSs. α, β, γ, δ and ε FSs are colored in green, blue, red, light blue and purple, respectively. (c) FS mapping in Z plane taken by p-polarization of $h\nu$ = 35 eV around the BZ center and circular-polarization of $h\nu$ = 40 eV around the BZ corner. X, Y in $k_X$ and $k_Y$ represents the tetragonal in-plane momentum axes. (d) Orbital characters on the hole and electron FSs obtained by first principle band calculations. The schematic FSs are based on ARPES results in (c). Intensity of α and β hole FSs is quite weak in (c) due to the matrix element effect. (e) $E$-$k$ image along cut 1 in (c) taken by circular-polarization at 60 K. (f) $E$-$k$ image along cut 2 in (c) taken by circular-polarization at 60 K. (g) $E$-$k$ image along cut 3 in (c) taken by p-polarization at 30 K. (h) $E$-$k$ image along cut 4 in (c) taken by circular-polarization at 30 K. Insets show the schematic band dispersions illustrating each $E$-$k$ image. Dotted curves represent the band dispersions whose intensity is suppressed due to the matrix element effect. $k_F$ for each band is indicated by the white triangle.

Laser-ARPES measurements were performed on a spectrometer built using a VG-Scienta R4000WAL electron analyzer and a VUV-laser of 6.994 eV [44] as a photon source at ISSP. Using the λ/2 (half-wave) plate, we can rotate the light polarization vector and obtain $s$- or $p$-polarized light without changing the optical path. The energy resolution was set to ~5 meV to get high count rate. ARPES measurements at the Brllouin zone (BZ) corner were performed using a spectrometer VG-Scienta R4000WAL electron analyzer, motor-controlled 6-axis manipulator and a Helium discharge lamp of $h\nu$ = 40.8 eV at University of Tokyo. The energy resolution was set to ~10 meV. The spectra were reproducible over measurement cycles of 12 hours. Fermi level ($E_F$) of samples was referenced to that of gold film evaporated onto the sample substrate. The single crystals were cleaved at 200 K in ultra high vacuum of ~5 × 10$^{-11}$ Torr. We did not observe any difference in data even when the crystals were cleaved at the lowest $T$. Synchrotron-based ARPES measurements were carried out at BL 9A of Hiroshima Synchrotron Radiation Center (HiSOR) and BL 28A of Photon Factory (PF). At HiSOR BL 9A, a VG-Scienta R4000 analyzer and circularly-polarized light were used with the total energy resolution of ~8 meV. At PF BL-28A, a VG-Scienta SES-2002 analyzer and circularly- and p-polarized light were used with the total energy resolution of ~10 meV. The crystals were cleaved in situ at $T$ ~10 K in an ultra-high vacuum of ~5 × 10$^{-11}$ Torr.

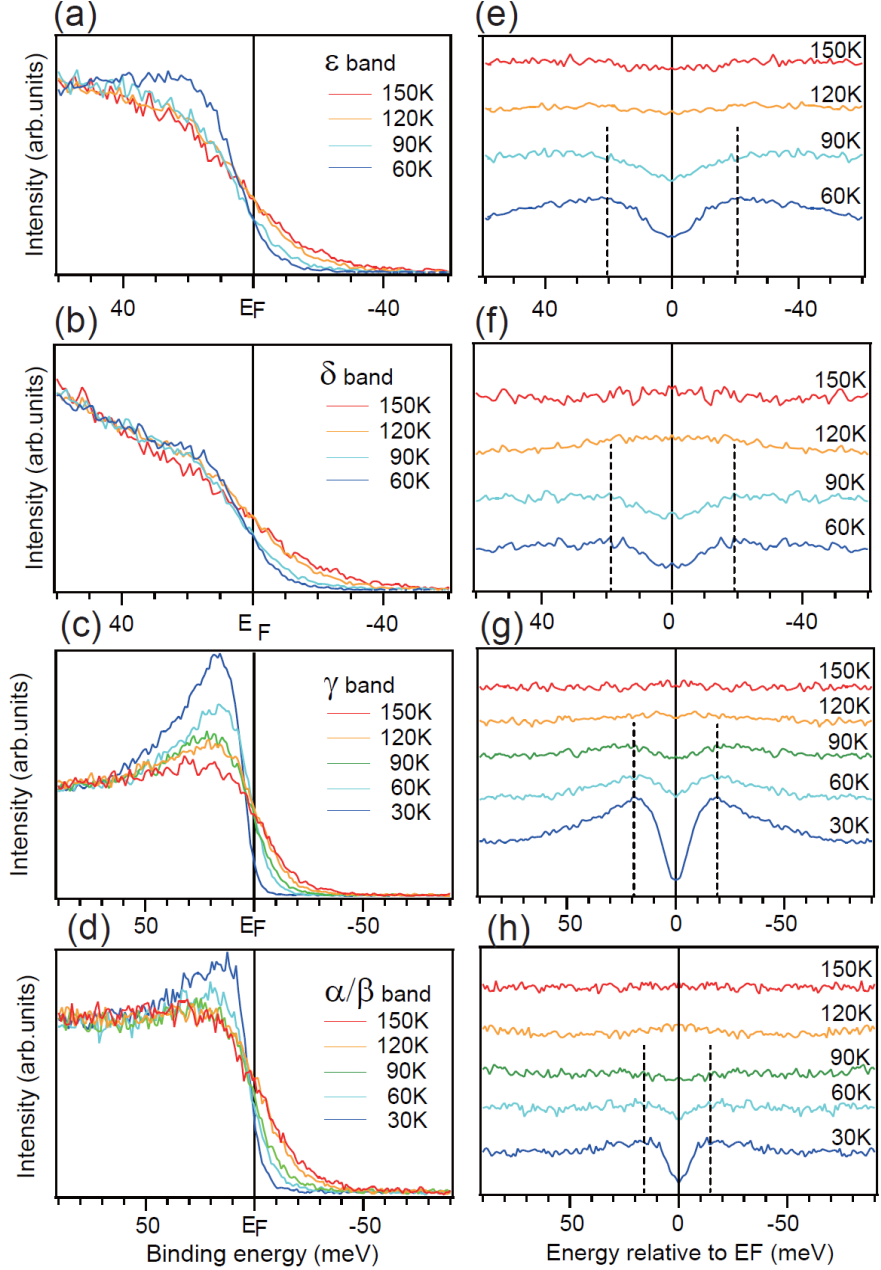

**Fig. 3.** (a-d) *T*-dependence of the EDCs for $x = 0.30$ measured at $k_F$ of the ε band in Fig.2 (e), δ band in Fig.2 (f), γ band in Fig.2 (g), α/β band in Fig.2 (h), respectively. (e-h) The EDCs in (a-d) symmetrized with respect to $E_F$ and further normalized by the smoothed EDC at 150 K, respectively. Broken lines indicate the PG energy. Note the larger x-axis energy scale for γ and α/β bands (panels (c,g) and (d,h)).

AsP122 system shows quasi-two dimensional FSs above $T_{N,s}$ as shown in Fig. 2(a). Three hole-FSs (α, β, γ) and two electron-FSs (δ, ε) exist around the BZ center and the BZ corner, respectively (Fig.2 (b)). Substituting As ions by isovalent P ion causes reduction of the pnictogen height without changing the carrier density. As P ion concentration is increased, the warping of the γ-hole FS along $k_z$ axis increases, resulting in the larger FS near Z plane [45]. Figure 2 (c) shows FS mapping in Z plane for $x = 0.30$ taken by using the synchrotron radiation photon source (PF BL-28A) of $h\nu = 35$ eV for the BZ center and $h\nu = 40$ eV for the BZ corner. As shown in Fig. 2 (d), the hole and electron FSs have different orbital characters depending on the momentum region.

In order to investigate the *T*-dependence of the fine electronic structure in the multi-orbital system, it is required to clearly separate the multiple band dispersions. For this purpose, we chose p- polarization for the observation of γ band, and circular polarization for α, β, δ and ε bands in the experimental geometries shown in Figure 2 (c). In Fig. 2 (e-h), we show the α/β (nearly degenerate), γ, δ and ε bands taken along cut 1 – 4 in Fig. 2 (c), respectively. Fermi momentum ($k_F$) was determined at each $T$ by fitting the momentum distribution curves (MDCs) near $E_F$ with a Lorentz function. Note that the energy distribution curves (EDCs) taken at $k_F$ indicated by the white triangles can be well separated from the intensity of other bands, in the vicinity of $E_F$.

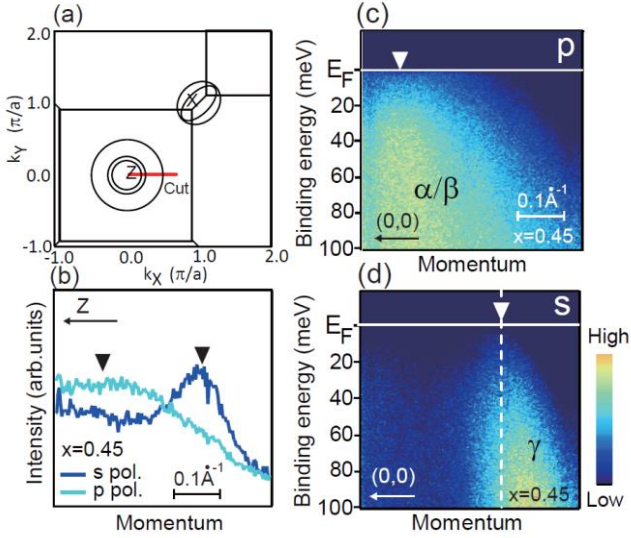

**Fig.4.** (a) Schematics of the in-plane cut of FS sheets in Z plane of BaFe$_2$(As,P)$_2$. Red line indicates the momentum region measured by laser-ARPES for $0.07 \leq x \leq 0.61$. Only for $x = 0.00$, the measured momentum cut is rotated by 45 degrees. (b) MDCs at $E_F$ for $x = 0.45$ recorded by s- (blue) and p- (light blue) polarizations, along momentum cut shown in (a). Black triangles indicates the $k_F$ for α/β and γ hole bands, determined by the peak positions of the MDCs. α and β hole bands are nearly degenerate. (c,d) Hole bands at $x = 0.45$ measured along the momentum cut shown in (a) by using p- and s-polarizations, respectively. γ hole band can be separately observed by choosing s-polarization in the experimental geometry shown in (a). The energy distribution curves in γ hole band were obtained at the $k_F$ as indicated by broken white line in (d).

$T$-dependences of the EDCs at $k_F$ for ε, δ, γ and α/β bands at $x = 0.30$ are displayed in Figure 3 (a-d). The coherence peak near $E_F$ is clearly observed at low $T$ especially for the hole bands. If we take a close look at $E_F$, all of the bands show the depression of the spectral weight with lowering $T$. In order to extract the intrinsic $T$-dependent part of the spectral weight, the EDCs are symmetrized with respect to $E_F$, and further normalized by the smoothed EDC recorded at the highest $T$ (150 K) as shown in figure 3 (e-h). Normalized EDCs in all bands indicate the suppression of the spectral weight near $E_F$ roughly below ~120 K. Since this gap-like structures concomitantly evolve both in the electron and hole bands, the $T$-dependent gapped feature in the EDCs in the normal state should be then attributed to PG formation in the total density of states. We thus consider that the hole and electron bands show the nearly comparable PGs of ~ 20 meV for optimally-doped AsP122.

In order to investigate how the PG evolves in the phase diagram, here we use laser-ARPES ($h\nu$ = 6.994 eV) and detect the electronic structure of γ hole band near Z plane (3Z$^2$-R$^2$ orbital) for $x = 0.00 – 0.61$ (Fig. 4(a)) [46, 47]. By choosing the s-polarized laser as the photon source, we can successfully separate the γ hole band from other bands (Fig. 4(b-d)), owing to the distinct orbital characters of the hole FS sheets [48] and selection rules for ARPES [49]. Here we focus on the γ hole band for investigating its $T$-dependence at $x = 0.00$ ($T_{N,s}$ = 136 K), 0.07 ($T_{N,s}$ = 114 K) 0.24 ($T_{N,s}$ = 55 K, $T_c$ = 16 K), 0.30 ($T_c$ = 30 K), 0.45 ($T_c$ = 22 K) and 0.61 ($T_c$ = 9 K) of AsP122.

Figure 5(a-c) show the $T$-dependence of the EDCs at the $k_F$ in the γ hole band at $x = 0.07$, 0.30 and 0.61. EDCs at $x = 0.61$ clearly intersect with each other at $E_F$, as simply expected by the $T$-dependence of the Fermi-Dirac (FD) distribution (Fig. 5(c)). In contrast, EDCs at $x = 0.07$ in Fig. 5(a) exhibit drastic $T$-dependence below ~150 K. As shown in the inset, the spectral weight at $E_F$ decreases with lowering $T$, indicating the evolution of a gap-like structure. Similar $T$-dependences are also observed for $x = 0.30$ (Fig. 5(b)), although the gap-like structure itself becomes weaker with increasing $x$. This result indicates that the gap-like structure exists in a wide $T$- and $x$-region, which seems to disappear toward $x \sim 0.6$.

The EDCs are symmetrized with respect to $E_F$, and further normalized by the smoothed EDC recorded at the highest $T$ (200, 170, 180 K for $x = 0.07$, 0.30 and 0.61), as shown in Fig. 5(d-f). The symmetrized EDCs for $x = 0.61$ again indicates $T$-independent spectral weight similar to normal metal, thus showing the accuracy and the validity of the data analysis. The symmetrized EDCs at 0.07 and 0.30 clearly show the depression of the spectral weight in the normal state, consistent with the PG formation in the synchrotron-based ARPES (Figure 3). However, the coherence peak of $\gamma$ hole band is not apparent in the case of laser-ARPES, which may be possibly due to the matrix element effect.

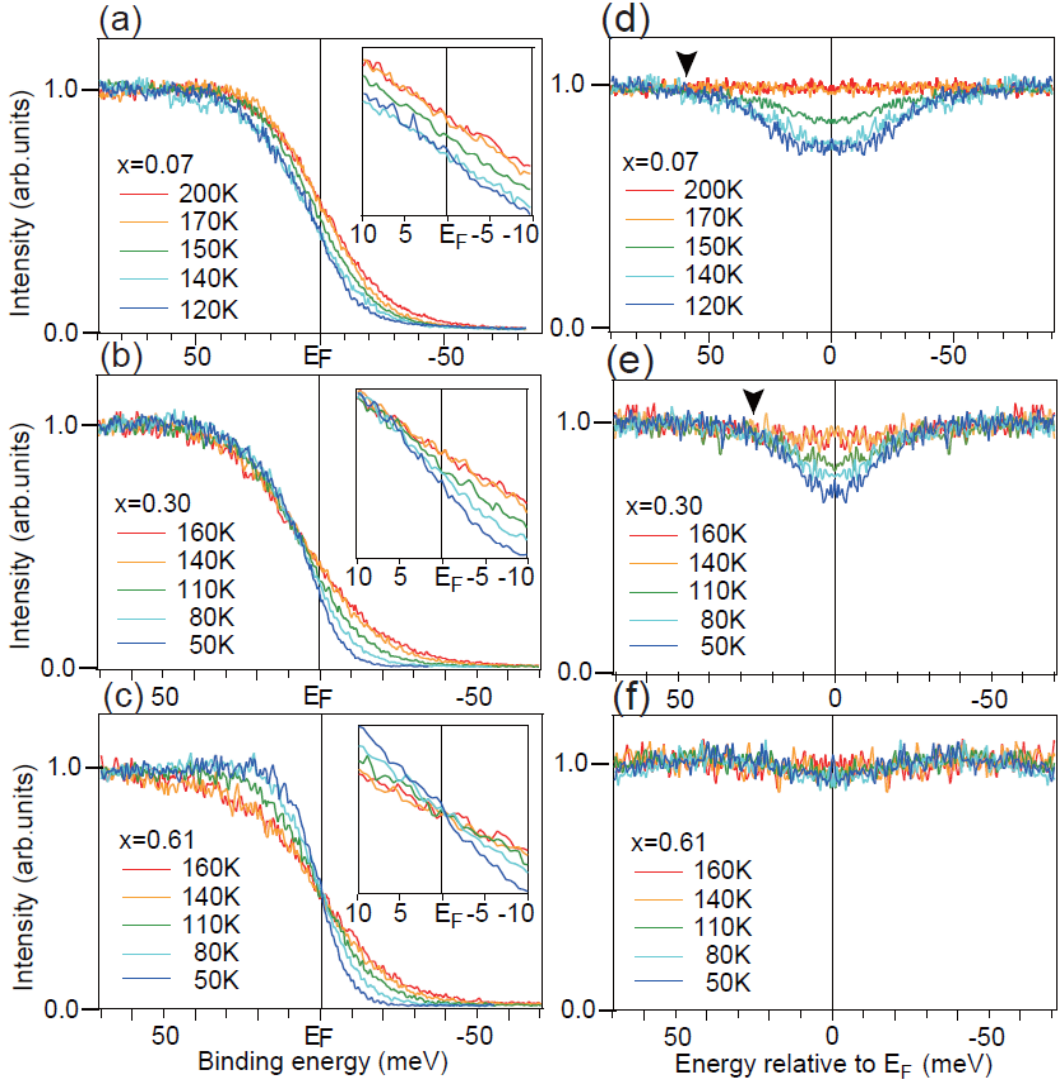

**Fig. 5.** (a-c) $T$-dependence of the EDCs at the $k_F$ of the $\gamma$ hole band measured for $x = 0.07$, 0.30 and 0.61, respectively. Inset shows the spectra near the $E_F$ in an enlarged energy scale. (d-f) EDCs symmetrized with respect to $E_F$ and further normalized by the smoothed EDC at the highest-$T$. Black arrow indicates the crossing point of the spectra with decreasing $T$, representing the PG energy. Note the larger x-axis energy scale for $x = 0.07$ (panels (a) and (d)).

The PG appears around a certain temperature $T^*_{PG}$. For example, at $x = 0.07$, the symmetrized EDC at $T = 150$ K slightly shows a gap feature near $E_F$, while those at $T = 170$ K and 200 K remain unchanged. $T^*_{PG}$ is thus estimated to be between 150 K and 170 K. Here we quantitatively estimate $T^*_{PG}$ for each P concentration. Figure 6(a-f) show the symmetrized EDCs (same as the spectra in Fig. 5(d-f)), with offsets and zero lines (horizontal black line) superimposed on respective curves as references, to focus on the $T$-dependent part of the spectra. The black area in each spectrum indicates the decrease of the spectral weight with lowering $T$, which should represent the PG formation. Here we note that previous ARPES reported the anomalous band shift with increasing $T$ toward room temperature on Ru-doped Ba122 [50] and Co-doped Ba122 systems [51]. In the present work, we are discussing the $T$-dependent spectral weight near $E_F$ separately from such band shift effect, by closely tracking the $T$-dependent band dispersion crossing $E_F$. We also note that the temperature ranges of the PG formation ($T^*_{PG} \leq 180$ K) and the band shift (up to 300 K) are not identical. This fact seems to indicate that these two phenomena are different in origin, which should be further investigated for understanding the anomalous T-dependent electronic structures.

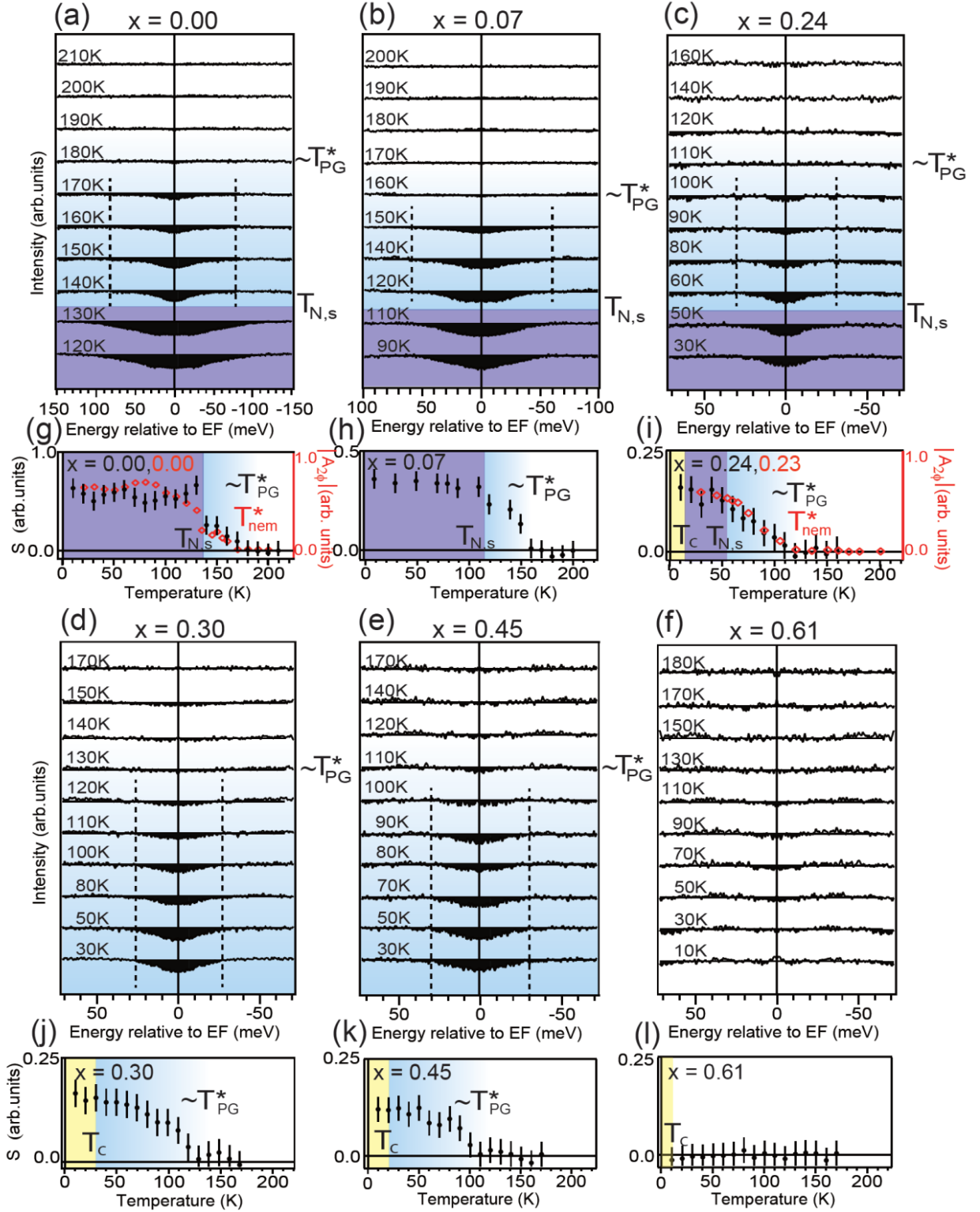

**Fig. 6.** (a-f) Symmetrized and normalized EDCs for $x$ = 0.00, 0.07. 0.24, 0.30, 0.45 and 0.61, respectively. The *T*-dependent depression of the spectral weight is indicated by the black shaded area. Note the larger x-axis energy scale for $x$ = 0.0 and 0.07 (panels (a) and (b)). Dotted lines indicate PG energy. (g-l) *T*-dependence of S, the gapped area. $T^*_{PG}$ is defined by the temperature at which S starts to evolve. Two-fold oscillation amplitudes of the torque for $x$ = 0.00 and 0.23 [27] are superimposed in (g) and (i), respectively. Background of each panel is colored in light blue for $T_{N,s} < T < T^*_{PG}$, blue for $T_c < T < T_{N,s}$ and yellow for $T < T_c$.

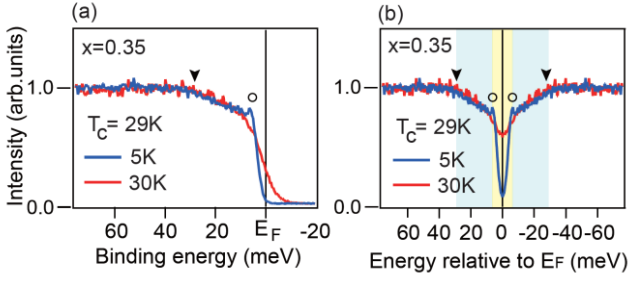

**Fig. 7.** (a) The EDCs of γ hole band measured at 5 K ($T < T_c$) and 30 K ($T_c < T$) for $x = 0.35$ by laser-ARPES. Black arrow and open circle represent the energy position of PG and SC gap, respectively. (b) The EDCs symmetrized with respect to $E_F$. Light blue and yellow area highlight the energy scale of the PG and SC gap, respectively.

Now we evaluate the *T*- and *x*-dependence of the PG. It is clear that the PG persists up to $T^*_{PG}$ ~ 180 K for $x = 0.00$ ($T_{N,s} = 136$ K), ~ 160 K for 0.07 ($T_{N,s} = 114$ K) and ~ 110 K for 0.24 ($T_{N,s} = 55$ K), as indicated by the spectra in the light blue area of Fig. 6(a-c). This is indicative of a PG phase extending much above $T_{N,s}$. The PG is still observed for further doped $x = 0.3$ and 0.45 where the magneto-structural transition no longer exists, but eventually disappears at $x = 0.61$, at least above 10 K. In Fig. 6(g-l), we plot the *T*-dependence of the gapped area S as a measure of PG formation, which is estimated by integrating the black area in Fig. 6(a-f). $T^*_{PG}$ estimated from the onset *T* of the PG formation are plotted onto the phase diagram in Fig. 1 (a). The obtained phase diagram of AsP122 clearly shows a PG phase extending above $T_{N,s}$ and $T_c$ in a wide range of $0.00 \leq x < 0.61$.

Recent torque magnetometry and X-ray diffraction measurements reported that two-fold symmetric properties in the lattice and magnetic response appear at $T^*_{Nem}$ (> $T_{N,s}$) [27]. Two-fold oscillation amplitudes of the torque for $x = 0.00$ and 0.23 are superimposed on Fig. 6(g) and 6(i), respectively. The good correspondence between $T^*_{Nem}$ and $T^*_{PG}$ in Fig. 6(g) and 6(i) reveals that such electronic nematic phase can be closely related to the PG observed by laser-ARPES, especially in the under-doped region.

The energy scale of the PG, $\Delta_{PG}$, can be also estimated from the crossing point of the symmetrized EDCs as indicated by the black arrows in Fig. 5(d) and (e). As summarized in the Figure 1 (b), it indicates that the PG tends to decrease from $\Delta_{PG}$ ~ 60 meV to $\Delta_{PG}$ ~ 25 meV with increasing P concentration from $x = 0.07$ to $x = 0.30$. Here we discuss the energy scales of the PG and SC gaps in AsP122 system. EDCs in the γ hole band below and above $T_c$ for $x = 0.35$ are shown in Fig. 7(a). Both EDCs show broad hump structure around ~ 30 meV (black arrow), indicative of the PG formation. In addition, the SC gap appears around ~ 5 meV with small coherent peak (open circle) in the EDC below $T_c$. Symmetrization of the EDCs in Fig. 7(b) clearly demonstrates that the PG energy is much larger than SC gap magnitude in AsP122.

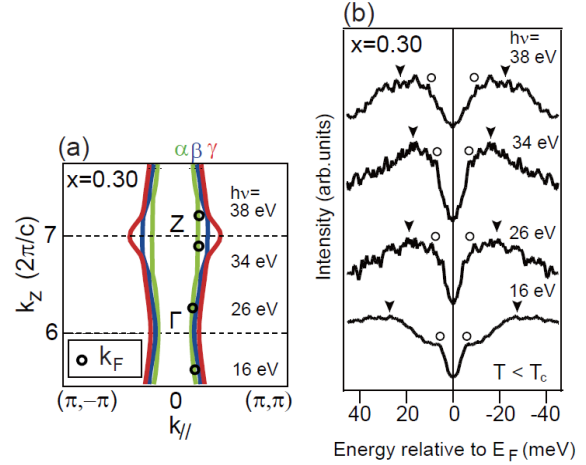

**Fig. 8.** (a) Schematics of three hole FSs obtained by first principle band calculations for $x = 0.30$. $k_F$ positions in α/β hole FSs at which the data in (b) were obtained are plotted by the open circles. (b) Symmetrized EDCs taken at $k_F$ in α/β hole FSs for $x = 0.30$ below $T_c$ by using photons of $h\nu = 16$ eV to 38 eV. Black arrows and open circles represent the energy scale of the PG and SC gap, respectively.

Synchrotron-based *hν*-dependent ARPES reveals that the PG structure in the EDCs are similarly found in the α/β hole FS around the BZ center. $k_F$ positions measured by photons of $h\nu = 16 - 38$ eV (HiSOR BL-9A and PF BL-28A) are superimposed on the calculated FSs in Fig. 8(a). Symmetrized EDCs shown in Fig.8 (b) were taken at $k_F$ of different $k_z$ values in α/β hole FS for $x = 0.30$ below $T_c$. Note that the spectral intensity of the EDCs of α/β hole band taken at several $k_F$ positions shown in Figure 8(a) is not affected by other bands, since these bands constitute the innermost FSs around the BZ center. The hump structures ranging from 18 meV to 30 meV (black arrows) are signatures of the PG formation. In particular, the shoulder structure around 5 - 8 meV (open circles) indicates SC gap opening, consistent with previous ARPES measurements [46,47,52]. Synchrotron-based ARPES thus show that SC gap in AsP122 is smaller than PG, consistent with laser-ARPES. PG with an energy scale higher than SC gap has been also reported for a variety of iron-based superconductors by several experimental probes [32,33,35,36,39,40].

Difference in the energy scales of the PG and SC gaps in several FSs imply that the PG formation is not simply a precursor of the SC electron pairing [4-6]. According to the $T^*_{PG}$ and $\Delta_{PG}$ values in Figure 1 (a) and (b), we can estimate the $2\Delta_{PG}/k_BT^*_{PG}$ as ~10.3 for $x = 0.00$, ~8.7 for $x = 0.07$, ~6.7 for $x = 0.24$, ~4.5 for $x = 0.30$ and ~6.3 for $x = 0.45$. Relatively high value for $x = 0.00$ and 0.07 implies importance of the magnetism for the PG formation. Indeed, $T$-dependence of $1/T_1T$ in NMR measurements suggests that the AF fluctuation is enhanced above $T_N$ in under-doped region [53,54].

One may consider that the energy scale of the PG reflects that of the AF gap below $T_N$. Previous ARPES [14,55] and dHvA [56] measurements on Ba122 reported the highly three-dimensional disconnected FSs in the AF state formed through the orbital-dependent band reconstructions across the magnetostructural transitions, as predicted by the first principle band calculations [14]. According to our $T$-dependent laser-ARPES on Ba122, the hole band we focused on in the present work intersects $E_F$ even in the AF state (not shown). It forms a part of the FSs and spectral intensity at $E_F$ is still finite, which is far from fully opened AF gap. While the scenario that the PG is a precursor of the AF gap sounds reasonable, it is not easy to apply to the present case.

It is also worth mentioning that the energy scale of the PG has the comparable order of magnitude as that of the anisotropic energy shift of zx and yz orbitals observed at the BZ corner. Such orbital inequivalency in zx/yz orbital induces the rotational symmetry breaking which is one of the possible origins of the electronic nematicity. The lifting of yz and zx orbital degeneracy in the band structure at the tetragonal BZ corner had been discussed in a previous ARPES study on Ba(Fe,Co)$_2$As$_2$ (FeCo122) [15]. Owing to the detwinning technique and polarization-dependent measurements, they separately observed the two bands below $E_F$ at (π,-π) point (Y point) and (π,π) point (X opint), indicating zx and yz character, respectively [15]. Such inequivalent shift in the energy of zx/yz orbitals were observed even above $T_{N,s}$, implying the development of the zx/yz orbital order from high $T$. The recent X-ray absorption measurements on FeCo122 system also reported the differences in occupation of zx/yz orbitals above $T_{N,s}$. [30].

Here we directly confirm the orbital inequivalency in AsP122 in relation to the existence of PG phase. Since the detwinning by uniaxial pressure is known to raise the onset $T$ of the anisotropy [15,57], here we chose twinned crystals to precisely compare with the phase diagram of the PG. Previous ARPES study confirmed that AsP122 has two electron bands (δ,ε) forming FSs around the BZ corner [45]. Along the momentum cut in Fig. 9(a), photoelectoron intensities of those electron bands are relatively weak due to the matrix element effect. We thus chose this experimental geometry in order to emphasize the zx/yz hole bands just below $E_F$ around the BZ corner.

Here we briefly describe how the orbital anisotropy can be observed in the twinned samples by referring to the reported case of Ba122 parent material [15]. The energy positions of zx and yz bands in schematics shown in Fig. 9 (b-e) roughly correspond to our ARPES results on P-substituted Ba122 system. Fig. 9 (b) and (c) show the schematic band dispersions at tetragonal phase for detwinned and twinned crystals, respectively. Here, zx and yz orbitals are degenerate reflecting the D4h symmetry. Therefore, there is naturally no difference in the $E$-$k$ image between detwinned and twinned crystals. On the other hand, Fig. 9 (d) and (e) show schematic band dispersions with orbital-dependent energy shift below $T^*_O$ for detwinned and twinned crystals, respectively. $T^*_O$ represents the temperature at which zx/yz bands start to show the inequivalent energy shift. Below $T^*_O$, zx orbital around Y point and yz orbital around X point show different energy shift (Fig. 9 (d)), as reported in the case for Ba122 [15]. In the case of twinned crystals, X and Y points are overlapped in the momentum-space, thus resulting in both the zx and yz bands with different energy levels appearing in one momentum cut across the BZ corner (Fig. 9 (e)). Then, the difference in energy between these two hole bands represents the anisotropic electronic occupation in zx/yz orbitals.

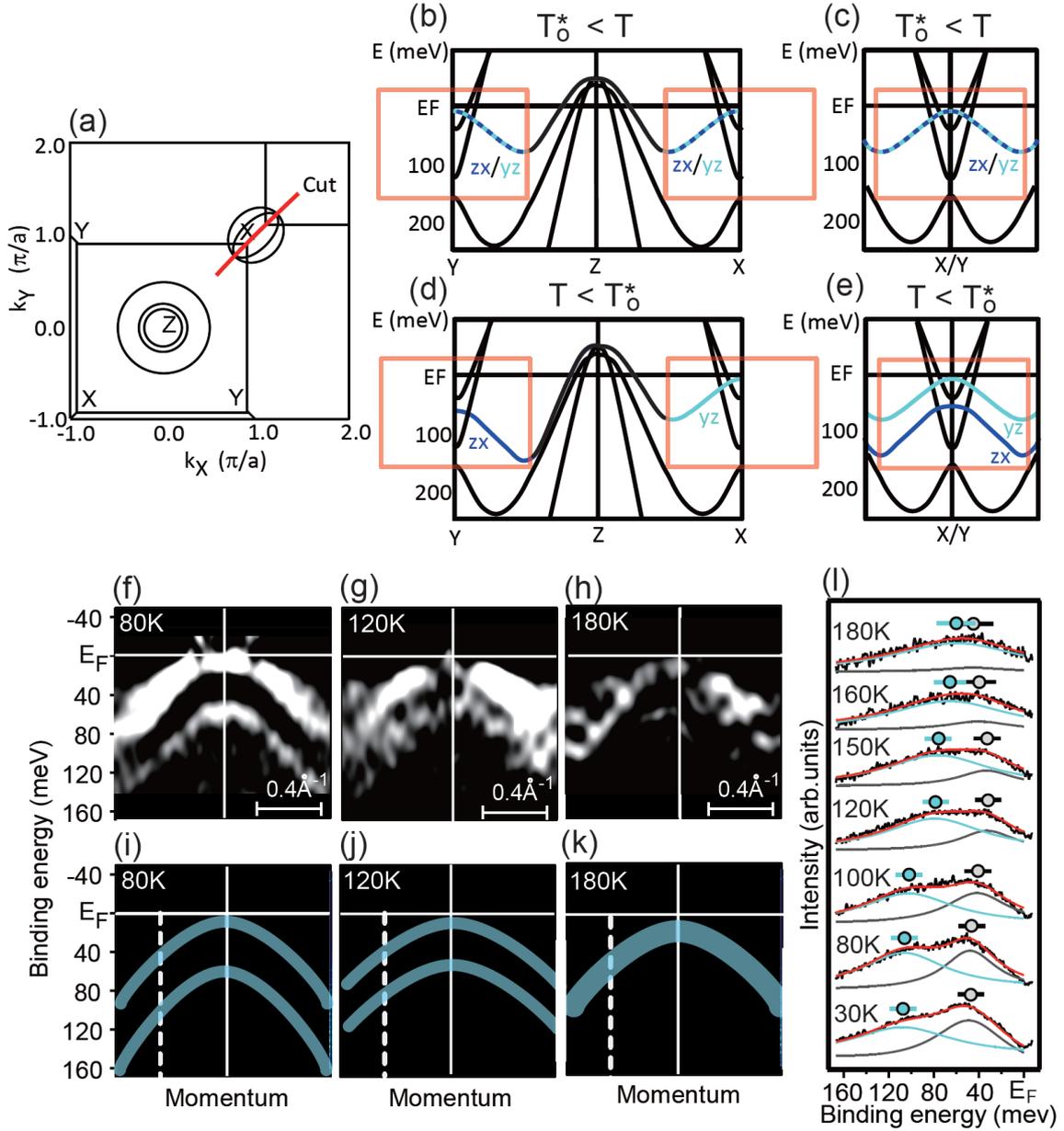

**Fig. 9.** (a) Schematics of the in-plane cut of FS sheets in Z plane of BaFe$_2$(As,P)$_2$. Red line indicates the momentum region measured by helium-lamp ARPES for $x = 0.07$. (b,c) Schematic band dispersions above $T^*_O$ for detwinned and twinned crystals, respectively, where $T^*_O$ represents the temperature at which zx/yz bands start to show an inequivalent energy shift. Red rectangles represent $E$-$k$ region measured by helium-lamp ARPES. (d,e) Schematic band dispersions with orbital-dependent energy shift below $T^*_O$ for detwinned and twinned crystals, respectively. This pictures was originally shown in Ref. 15 for Ba122 parent material. Energy positions of zx and yz bands in (b-e) are modified taking into account our ARPES results on P-substituted Ba122 system. (f) Band dispersion (second derivative) at the BZ corner near Z plane measured by the photons of $h\nu = 40.8$ eV at 80 K ($T < T_{N,s}$). (g) The same measured at 120 K ($T_{N,s} < T < T^*_{PG}$). (h) The same measured at 180 K ($T^*_{PG} < T$). (i-k) Schematics of the band dispersions corresponding to (f-h), respectively. Two hole-bands (light blue curves) are observed at low-$T$. Note that intensity of δ and ε electron bands crossing $E_F$ is weak in this experimental geometry. (l) $T$-dependence of the EDCs divided by FD function at the momentum cut indicated by the broken white line in (i-k). Linear back ground was subtracted from each spectrum. Red curves represent the fitting functions composed of double Lorentz functions illustrated by gray and light blue curves. Gray and light blue circles represent the peak positions in the EDCs obtained by the fitting procedure, indicating the energy position of upper and lower hole band, respectively.

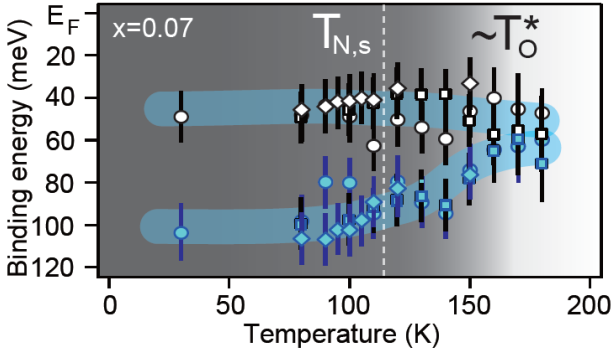

**Fig. 10.** *T*-dependence of the energy position of the two hole bands below $E_F$ around the BZ corner. Gray and light blue symbols represent the peak positions in the EDCs along a white broken line in (i-k) for upper and lower hole-bands, respectively. Circles, diamonds and squares represent the data obtained from different samples. $T^*_O$ is estimated to be 160 K ~180 K.

Shown in Fig. 9(f-h) are the band dispersions (second derivative) at the tetragonal BZ corner near Z plane taken along the momentum cut indicated in Fig.9(a) for $x = 0.07$ measured at 80 K, 120 K and 180 K, respectively. Corresponding *E-k* region is represented by the red rectangles in Fig.9 (b-e). Two hole bands are clearly observed below $E_F$ in the ARPES image at 80 K ($T < T_{N,s}$) in Fig. 9(f) (see also Fig. 9(i)). Note that both the zx/yz hole bands below $E_F$ at X and Y point appear in the same image as described in Fig.9 (e). Importantly, the hole bands remain well separated even at 120 K ($T_{N,s} < T < T^*_{PG}$) (Fig. 9(g) and (j)). At 180 K ($T^*_{PG} < T$), such a clear separation is not observed, indicating that the two bands are degenerate within experimental accuracy (see Fig. 9(h) and (k)).

Focusing on the energy positions of the two hole bands, we investigated the *T*-dependence of the EDCs at a certain momentum indicated by the broken white lines in Fig. 9(i-k). We plotted the EDC peak positions from 30 K to 180 K by fitting them using double Lorentz functions as shown in Fig. 9(l). EDCs were divided by FD functions and linear back ground was subtracted from each spectrum. Two peaks are marked by black (upper hole-band) and light blue (lower hole-band) circles. Difference in energy between upper and lower hole-bands decreases as *T* increases as summarized in Fig. 10. While two hole-bands are well separated at low *T*, they become nearly degenerate around 160 -180 K. The observation of the nearly degenerate bands splitting into two hole-bands at the BZ corner on cooling thus corresponds to the *T*-dependent inequivalent energy shift in zx/yz orbitals possibly related to orbital ordering, as was predicted theoretically [20-22]. We find that the inequivalent energy shift in zx/yz orbitals is similar in magnitude to the PG in $3Z^2$-$R^2$ orbital. It also sets in around $T^*_O$ ~ 160 K - 180 K, nearly comparable to $T^*_{PG}$. Although $T^*_O$ and $T^*_{PG}$ show large error bars due to the crossover-like behavior, these temperatures are close to $T^*_{Nem}$ [27]. This result implies that the rotational symmetry breaking in AsP122 occurs not only in the lattice and magnetism but also in the orbital, and may be related to the PG formation.

In this work, we confirmed PG formation in the momentum-resolved multi-band electronic structure of optimally-doped AsP122. Hole FSs around the BZ center and electron FSs near the BZ corner in Z plane commonly exhibit PG formation with a comparable energy scale of ~20 meV. The energy scale of the PG is larger than that of SC gap, suggesting that it is not a precursor of the SC pairing. The present results that the PG exists in all bands in Z plane (figure 3) and at any $k_z$ position in the hole bands (figure 8) suggest that the PG should appear in the total density of states as well.

We further found the PG phase well above $T_{N,s}$ and $T_c$. The shape of the PG phase diagram looks like that of the AF phase extended in a wider *T*- and *x*-region, especially in the under-doped region, as is the case for the cuprates. It implies that the AF fluctuation should be a key ingredient for PG formation. In cuprates, ARPES measurements reported the PG formation temperature above the SC dome [58]. There it has been suggested that the appearance of the electronic nematicity is associated with the PG formation by neutron scattering [59], STM [60] and Nernst effect [6] measurements. Therefore our results indicate that the PG phase and the electronic nematicity appear to be common features of high-$T_c$ superconductivity, both in cuprates and iron-pnictides.

However, there is a crucial difference between the two systems. In AsP122, the inequivalent energy shift in zx/yz orbitals is observed below $T^*_O$ ~ $T^*_{PG}$ (and also ~ $T^*_{Nem}$ [27]) for $x = 0.07$ (Fig. 1(a)). This implies that the origin of the PG is likely to be linked to the orbital ordering, which may give rise to the inplane anisotropic spin and/or orbital fluctuations causing the PG formation. Similar situation has been also theoretically discussed for the nematic phase [61, 62]. Complex coupling of spin, orbital and lattice [22,63] for respective compounds should be taken into account for further quantitative comparison. We stress that such orbital-related mechanism is not applicable to cuprates composed of a single $d_{x2-y2}$-orbital.

It is notable that PG phase in AsP122 is robust over the SC phase. On the contrary, PG of $BaKFe_2As_2$ (BaK122) which was reported for under-doped $x = 0.25$ [41] has not been found in optimal doping so far. Although the PG phase in BaK122 is still not completely established, it may indicate that the PG phase of BaK122 is much more suppressed than AsP122.

Possible variety of the PG phase in $BaFe_2As_2$ (Ba122) family can be consistently classified in relation to the electronic nematic phase. Actually, nearly isotropic inplane electronic resistivity is reported in under-doped BaK122 [64-66], indicative of the small contribution of the electronic nematicity. At least in Ba122 family, a close relationship between PG formation and electronic nematicity is expected although it seems to strongly depend on the character of the substitution ions. Understanding how the PG formation and orbital inequivalency reported here influence the normal state physical properties, which in turn should affect the variety of the SC gap symmetry in iron-based superconductors, is therefore a pressing issue, which deserves comprehensive studies of high-$T_c$ superconductivity.

**Acknowledgement:** We acknowledge H. Kontani, S. Onari, M. Ogata, K. Ishida and R. Arita, for valuable discussions. We thank T. Kiss for experimental support. Synchrotron-based ARPES experiments were carried out at BL-9B at HiSOR (Proposal No. 10-B-27 and 11-B-1) and BL-28A at Photon Factory (Proposal No. 2009S2-005, No. 2012S2-001, No. 2012G075 and No. 2012G751).

* Correspondence and requests for materials should be addressed to T.S. (shimojima@ap.t.u-tokyo.ac.jp)